\documentclass{mem}
\usepackage{natbib}\usepackage{txfonts}\usepackage{balance}
\usepackage{graphicx}
\usepackage[a4paper]{hyperref}
\idline{75}{282}

\newcommand{\iso}[2]{\hbox{${}^{#1}{\rm #2}$}}
\newcommand{\Msun}{\ensuremath{{\rm M}_{\sun}}}

\begin{document}

\title{Helium enhancements in globular cluster stars
from Asymptotic Giant Branch star pollution
}

   \subtitle{}

\author{
A. \,Karakas\inst{1}, Y. Fenner\inst{2}, Alison Sills\inst{1}, S. W. Campbell\inst{3}
\and J. C. \, Lattanzio\inst{3}
          }

  \offprints{A. Karakas}

\institute{
Origins Institute and Department of Physics \& Astronomy,
McMaster University, 1280 Main St. W., Hamilton ON L8S 4M1, Canada
\and
Harvard-Smithsonian Center for Astrophysics, 60 Garden Street,
     Cambridge MA 02138 USA
\and
Centre for Stellar \& Planetary Astrophysics, School of Mathematical Sciences
  PO Box 28M, Monash University, Clayton VIC 3800, Australia\\
\email{karakas@physics.mcmaster.ca}
}

\authorrunning{Karakas et al. }

\titlerunning{He in GC stars from AGB pollution}

\abstract{
Using a chemical evolution model we investigate the intriguing suggestion
that there are populations of stars in some globular clusters
(e.g. NGC 2808, $\omega$ Centauri) with enhanced levels of helium
($Y \sim$ 0.28 to 0.40)
compared to the majority of the population that presumably have a
primordial helium abundance.  We assume that a previous generation of
massive low-metallicity Asymptotic Giant Branch (AGB) stars has polluted
the cluster gas via a slow stellar wind. We use two independent sets of
AGB yields computed from detailed models to follow the evolution of
helium, carbon, nitrogen and oxygen in the cluster gas using a
Salpeter initial mass function (IMF) and a number of top-heavy
IMFs.  In no case were we able to fit the observational
constraints, $Y > 0.30$ and C$+$N$+$O $\approx$ constant.
Depending on the shape of the IMF and the yields, we either 
obtained $Y \gtrsim 0.30$ and large increases in C$+$N$+$O 
or $Y < 0.30$ and C$+$N$+$O $\approx$ constant.
These results suggest that either AGB stars alone
are not responsible for the large helium enrichment or that any dredge-up
from this generation of stars was less than predicted by standard models.
}
\maketitle{}

\section{Introduction}
Understanding the history and evolution of galactic globular cluster
(GC) stars poses one of the greatest challenges to astrophysics.
The star-to-star abundance variations of the light
elements C, N, O, Na, Mg and Al observed in every well studied cluster
to date \citep[][and references therein]{kraft94,gratton04}
are not found in field stars of the same metallicity \citep{gratton00}.
Hence these abundance anomalies are somehow the result of the cluster
environment. The variations of the elements follow a common pattern
from cluster to cluster:  C-N,  O-Na and Mg-Al are all negatively
correlated \citep{shetrone96a,kraft97,cannon98,gratton01,cohen05a,cohen05b}.
The abundances of iron-peak, s and r-process elements do not show the
same star-to-star scatter as the light elements nor do these
elements vary {\em with} the light elements \citep{gratton04,james04,yong05c}.
The key points are that O has been destroyed in some stars by up to one
order of magnitude, the C$+$N$+$O and Mg$+$Al abundances remain almost
constant  regardless of the absolute spread and there is no evidence
for large-scale variation of neutron-capture elements, except
in the case of $\omega$ Cen \citep{smith00}.
\par
The self-pollution scenario, first proposed by \citet{cottrell81},
is the most promising to explain these abundance trends.
This is because the  star-to-star abundance 
variations in C, N, O and Na are observed in stars at 
or near the main-sequence turn-off in addition to stars along
the giant branch
\citep{gratton01,ramirez03,james04,cohen05a,cohen05b}.
The other scenario, deep mixing, is still required to operate 
in low-mass giants to convert some C to N after the luminosity bump 
\citep{charbonnel94}.  Owing to the constant [Fe/H] 
of stars in a given GC it has been assumed that the source of the 
pollution was intermediate-mass Asymptotic Giant Branch (AGB) stars 
with initial masses between $\sim$3 to 8$\Msun$ rather than supernovae.
The hot bottom burning experienced by these objects provides an
environment (at least qualitatively) to produce helium, convert 
C and O to N, Ne to Na and Mg to Al \citep{lattanzio04}.
The mass lost via the slow winds of AGB stars could, in principle,
have been retained by the cluster from which new stars may have
been born \citep{thoul02}.
\par
There is an increasing amount of evidence for helium enrichment from
horizontal branch (HB) morphology and main-sequence colour-magnitude diagrams
\citep{norris04,dantona04,lee05,caloi05,piotto05} where the helium has
been proposed  to have come from a previous generation of low-metallicity
intermediate-mass AGB stars. For example, the unusual HB morphology of
NGC 2808, which exhibits an extended
blue tail and a gap separating the red and blue clumps \citep{bedin00},
can be best explained if the blue stars have an enhanced amount of helium
(up to $Y \sim 0.32$) compared to those in the red HB clump
with a primordial $Y \approx 0.24$ \citep{dantona04}.
Further evidence for an enrichment
in helium in NGC 2808 comes from the peculiar main sequence
\citep{dantona05}, where the bluer stars are inferred to have
$Y \approx 0.40$.  Observations by \citet{piotto05} showed that the 
blue main-sequence of $\omega$ Centauri 
is more metal-rich than the red sequence, contrary to what is expected 
from stellar evolution, and \citet{norris04} showed that isochrones 
with $Y=0.40$ fit the bluest stars. These intriguing pieces
of observational evidence have motivated us to study the AGB
self-pollution scenario from a global perspective.
\par
Here we use the \citet{fenner04} GC chemical evolution model to
follow the evolution of helium in the intracluster gas. We
explore AGB model uncertainties by using two independent sets
of AGB yields, including those used in the previous study which were
tailor made for NGC 6752 with a metallicity
[Fe/H] $\approx -1.4$. This metallicity is slightly more metal-rich
than the average metallicity of [Fe/H] $\sim -1.6$ for NGC 6752, 
$\omega$ Centauri, M3 and M13 with data taken from the catalogue 
of \citet{harris96}. The cluster NGC 2808 is more metal-rich 
with [Fe/H] $\sim -1.15$ \citep{harris96} but we feel these AGB yields
are suitable for this study because the yields would not change 
significantly in this [Fe/H] range.
We also follow the evolution of C, N and O since they impose
important empirical constraints, i.e. C$+$N$+$O $\approx$ constant
that must be met by the model.

\section{Helium Production in Asymptotic Giant Branch Stars} \label{sec:heprod}

\begin{table*}
\caption{The mass of helium (in $\Msun$) expelled into the intracluster medium}
\label{table1}
\begin{center}
\begin{tabular}{lccccccc}
\hline
\\
Model type & 1.25 & 2.5 & 3.5 & 5.0 & 5.5 & 6.5 \\
\hline \\
\citet{fenner04} & 0.160 & 0.490 & 0.680 & 1.53 & -- & 1.92 \\
\citet{ventura02} & -- & -- & 0.675 & 1.14 & 1.30 & -- \\
\\
\hline
\end{tabular}
\end{center}
\end{table*}

The helium yields from the AGB stars used by \citet{fenner04}
are shown in table~\ref{table1} as the total mass of \iso{4}He expelled into
the intracluster medium by each model. We hereafter refer to the yields 
used by \citet{fenner04} as {\em our yields}.  We also show the 
$Z=0.001$ yields from \citet{ventura02} for comparison and note that 
they agree to within $\sim 30$\%. 
Our yields \citep{fenner04} are systematically larger even
though \citet{ventura02} use a different convective model
and mass-loss rate and observe shallow dredge-up in their computations.
The relatively good agreement comes about because most of the helium
mixed to the surface in $m \gtrsim 3\Msun$ models comes from the 
second dredge-up, not the thermally-pulsing AGB phase. Also, helium
yields are more robust to reaction rate uncertainties than other
species (e.g. \iso{23}Na) owing to the fact that the net result of 
hydrogen burning is helium fusion, regardless of the internal rates 
of the various cycles (CNO, NeNa and MgAl).

\section{The initial mass function}

The globular cluster chemical evolution model was described in
detail in \citet{fenner04}. We made two main changes for this study:
we used an independent set of AGB yields from \citet{ventura02},
and we changed the shape of the initial mass function; see \citet{karakas06b}
for more details.  There are other AGB yields besides those we consider here
although most are from synthetic computations (e.g. \citet{izzard04})
and for more metal-rich populations. \citet{herwig04b} has low-metallicity AGB 
yields computed from detailed models but the results of study would not
change significantly if we were to have used these yields instead. 
This is because his results are somewhere between the  Ventura et al. 
and the Monash models. That is they have deep dredge-up but only a few 
(less than 20) thermal pulses, whereas the Monash models of 
$\sim 5\Msun$ presented in \citet{fenner04} have $\sim 100$ thermal 
pulses plus deep dredge-up.

In this section we focus on one
crucial element of the model -- the initial mass function (IMF). 
Most discussions in the literature point toward a universal mass
function \citep{kroupa01} although there are hints for variations
at low metallicity.  \citet{dantona04} comment 
that to produce the amount of helium required to form the number of 
blue HB stars requires a factor of $\sim 10$ more 4 to 7$\Msun$ stars than
produced by a standard Salpeter IMF.  Stars between 1 to 3$\Msun$
are either not produced or ejected from the
cluster owing to a lack of evidence for enhanced levels of s-process
elements or carbon. However, there is little physical motivation why such an
unusual IMF would result from the early evolution of GCs when there is
no observational evidence that such strange IMFs formed in other systems
of comparable mass such as dwarf spheroidal galaxies. Although the
abundance patterns observed in these systems are remarkably different from
those in (most) globular clusters \citep{pritzl05} there is still
no evidence for a top heavy IMF.

\begin{figure*}[t!]
\resizebox{\hsize}{!}{\includegraphics[clip=true]{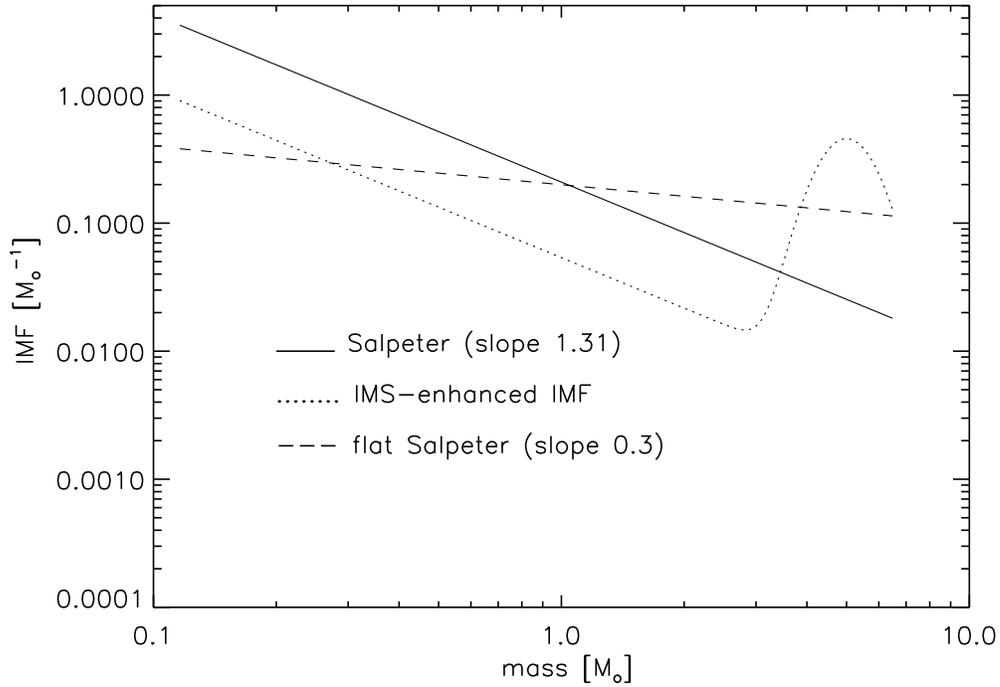}}
\caption{\footnotesize Possible initial mass functions for the first
generation of globular cluster stars. We consider three possible 
choices for the IMF including a Salpeter with slope $s = 1.31$, a 
flat Salpeter with $s=0.30$ and an IMF the IMS-enhanced IMF places 
about 10 times more mass in 3.5 to 6.5$\Msun$ stars.
}
\label{imf}
\end{figure*}

With these considerations in mind, we compute simulations with a
standard Salpeter-like IMF with a slope of 1.31, a flat Salpeter with 
slope of 0.30 and a top-heavy IMF that increases the
number of intermediate-mass AGBs by a factor of 10, see fig.~\ref{imf}.
We hereafter refer to the top-heavy IMF as ``IMS-enhanced''
(intermediate-mass star enhanced).

\section{Results}

\begin{figure*}[t!]
\begin{tabular}{cc}
\resizebox{0.45\hsize}{!}{\includegraphics[clip=true]{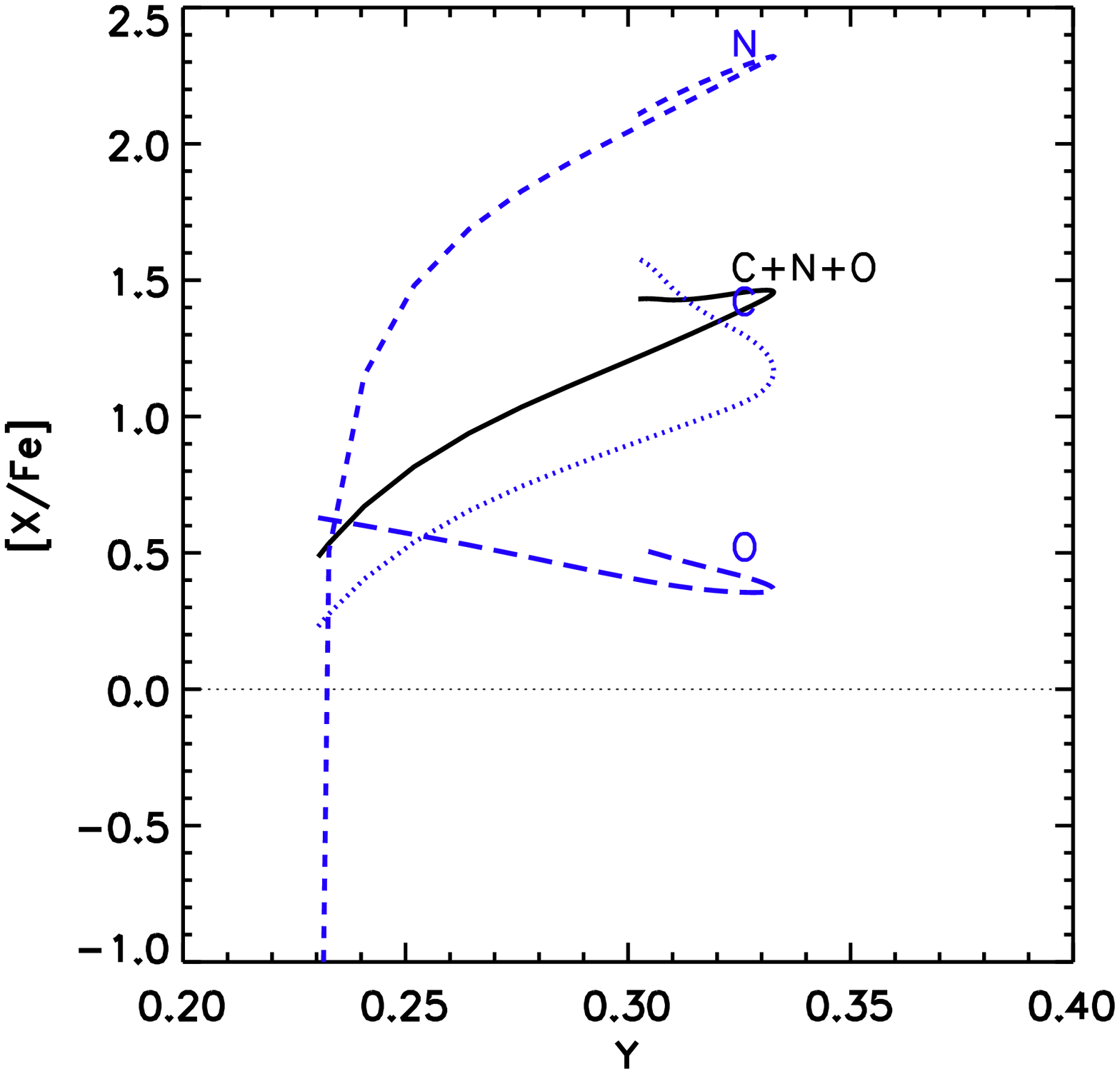}} &
\resizebox{0.45\hsize}{!}{\includegraphics[clip=true]{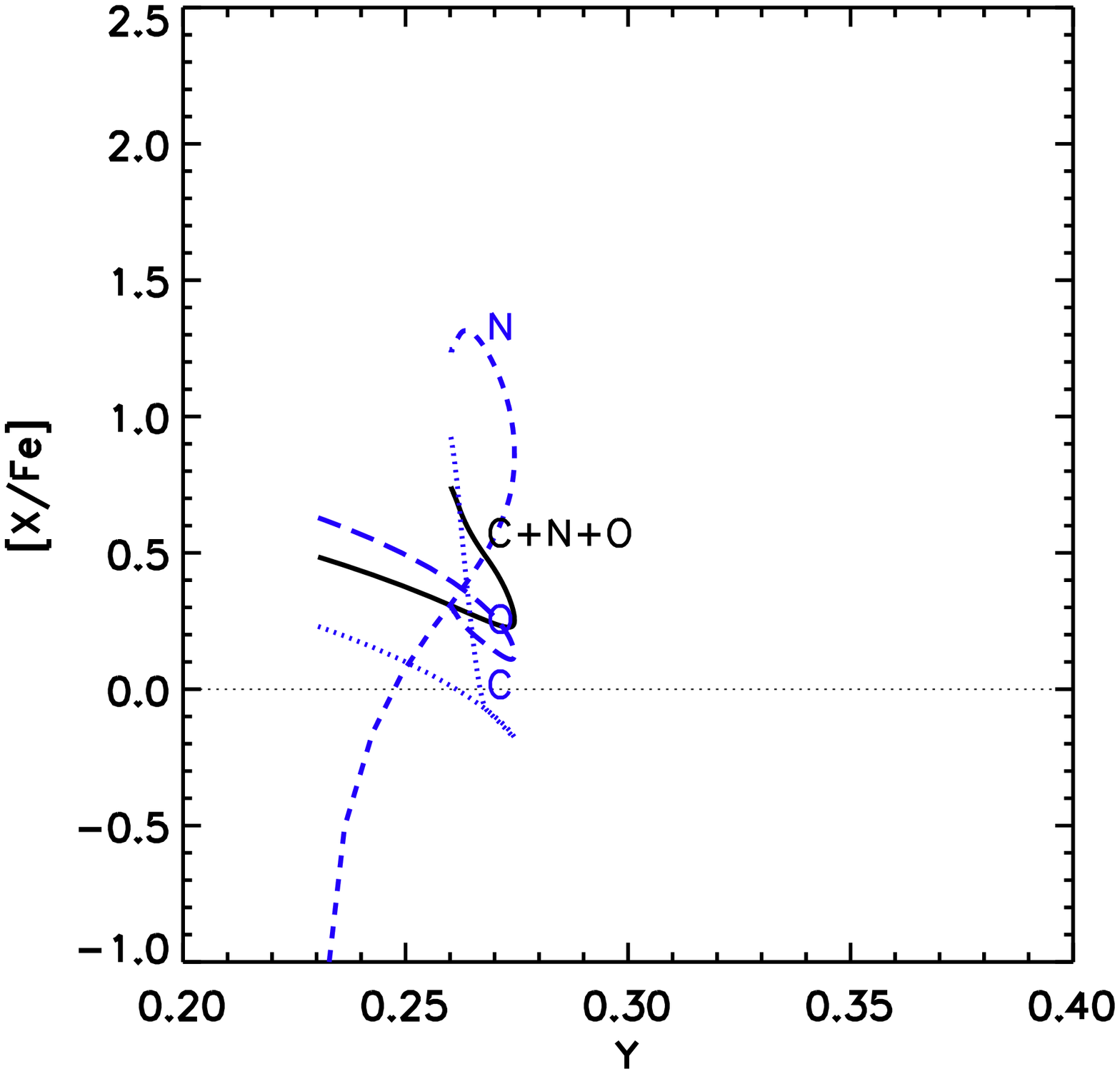}}
\end{tabular}
\caption{\footnotesize The temporal evolution of the C$+$N$+$O abundance ([CNO/Fe])
in the cluster gas as a function of $Y$ assuming a flat Salpeter IMF with slope $=0.30$.
In the ({\em left}) panel we show results using yields from \citet{fenner04}
and in the ({\em right}) panel yields from \citet{ventura02}.
}
\label{results}
\end{figure*}

In fig.~\ref{results} we show the temporal evolution of the C$+$N$+$O 
isotopes (on a $\log$ scale) as a function of the helium mass fraction
assuming the flat Salpeter IMF.   Using our yields we
see a substantial increase in helium with a maximum abundance
of $Y \approx 0.33$, similar to (but smaller than) the value required 
by isochrones 
\citep{dantona04,dantona05} to match the bluest HB and main-sequence 
stars of NGC 2808.  The maximum helium abundance is reached in under
200~Myr reflecting the dominant contribution of intermediate-mass AGB
stars with lifetimes $\lesssim 120$~Myr. The increase in helium is 
also accompanied by a $\sim 1$~dex increase in C$+$N$+$O although 
there is some depletion of O by $\sim 0.3$~dex.
\par
The simulation using the Ventura et al. yields has a moderate
increase in the total C$+$N$+$O of $\sim 0.5$~dex but this is probably
within observational errors.  The maximum $Y$ in this case
does not exceed 0.27, well below the inferred value of the bluest
main sequence and HB stars. However, the substantial O depletion
of $\sim 0.5$ dex in this case is similar to (but still smaller than)
the maximum dispersion observed in GC stars.  If we compare
to observations, the most ``polluted'' stars in M13 have
[O/Fe] $\approx -0.8$ \citep{kraft97} whereas ``normal'' stars
have [O/Fe] $+0.4$, indicating significant O destruction of
more than one order of magnitude. The simulation with the flat
Salpeter produces many more stars with $m > 1\Msun$ compared
to a standard Salpeter (see fig.~\ref{imf}) resulting in more
carbon along with helium injected into the intracluster medium.
Indeed, this IMF produces results quite different to the IMS-enhanced
IMF which gives higher weight to the most massive AGB stars which
have more efficient HBB and a lower weight to stars with
$1 \lesssim m (\Msun) \lesssim 3$.
Results using the standard Salpeter and the IMS-enhanced are
presented in \citet{karakas06b} and are also not consistent with
the observational constraints we have considered in this study.

\section{Discussion \& Conclusions}

Our investigation into the chemical evolution of helium in globular
clusters has shown the difficulty the AGB self-pollution scenario suffers
when trying to explain the large helium enrichment ($Y \gtrsim 0.30$) 
hypothesized to fit the horizontal branch morphology of clusters like NGC 2808.
We utilize a chemical evolution model and two independent sets of AGB 
yields to follow
the evolution of the intracluster medium for a typical cluster
with [Fe/H] $= -1.4$. We have tested three different IMFs for the
first generation of GC stars including a standard Salpeter, a flat
Salpeter and one IMF that boosts the number of intermediate-mass AGB
stars by a factor of 10. The flat Salpeter produces maximum helium
mass fractions in-between those found when using the standard
Salpeter (with much lower $Y$ values) and the IMS-enhanced IMF
with slightly larger $Y$ abundances \citep{karakas06b}.
The behaviour of the C, N and O elements is however quite
different, owing to the overall increased number of 1 to 6.5$\Msun$
stars which produce C as well as helium.

Simulations assuming a flat Salpeter with a slope of 0.30 show 
helium mass fractions as large as $Y \sim 0.33$ but only with 
enormous increases in  the total C$+$N$+$O content of the 
cluster gas, in violation of observations.  The Ventura et al. yields 
predict a maximum $Y \approx 0.27$ with the total C$+$N$+$O abundance 
increasing by $\sim 0.5$\,dex.  Assuming that such 
an IMF is realistic we still have a problem fitting all
of the observational constraints.  Indeed, the use of such
an IMF does not help the difficulties faced by the self-pollution
scenario in matching the constraints
that we have considered  in this study i.e. $Y \gtrsim 0.30$
and C$+$N$+$O $\approx$ constant. We note that the previous
study by \citet{fenner04} suffered similar difficulties
when trying to match the observed abundances of the Na, Mg and
Al isotopes in the cluster NGC 6752.

There are many model uncertainties and in particular the extent
of the dredge-up is far from known and shallower dredge-up, as
observed in the Ventura et al. models, would help keep
C$+$N$+$O constant. The nucleosynthesis associated with
hot bottom burning is also dependent on the convective
model (or the mixing-length parameter $\alpha$ for MLT models)
and varying this will impact the yields \citep{ventura05a}.
However, more efficient convection leads to larger luminosities
which would likely drive mass loss, leading to shorter
AGB lifetimes and possibly smaller helium yields.

Thus far we have assumed that the large helium
abundances inferred from theoretical isochrones are accurate.
These determinations are model dependent and suffer from
many of the same uncertainties that afflict AGB models
(e.g. convection) and hence it is hard to gauge just how
reliable these helium determinations are. Even if they are 
overestimates with the largest $Y$ closer to 0.30 instead
of 0.40, we still have a problem fitting the observational
constraints with the current set of AGB models. 
Perhaps given the difficulties associated with the self-pollution
scenario we need to look to other solutions including pollution
from outside the cluster, as discussed in the case of
$\omega$ Centauri by \citet{bekki06}.
 
\begin{acknowledgements}
AK thanks the organising committee for covering local expenses
and for organising a wonderful meeting in Granada!
\end{acknowledgements}

\end{document}